# Unsupervised Reward-Driven Image Segmentation in Automated Scanning Transmission Electron Microscopy Experiments

Kamyar Barakati[1, a], Utkarsh Pratiush[1], Austin C. Houston[1], Gerd Duscher[1], and Sergei V. Kalinin[1,2, b]

[1] *Department of Materials Science and Engineering, University of Tennessee, Knoxville, TN 37996, USA*
[2] *Pacific Northwest National Laboratory, Richland, WA 99354*

**Abstract**

Automated experiments in scanning transmission electron microscopy (STEM) require rapid image segmentation to optimize data representation for human interpretation, decision-making, site-selective spectroscopies, and atomic manipulation. Currently, segmentation tasks are typically performed using supervised machine learning methods, which require human-labeled data and are sensitive to out-of-distribution drift effects caused by changes in resolution, sampling, or beam shape. Here, we operationalize and benchmark a recently proposed reward-driven optimization workflow for on-the fly image analysis in STEM. This unsupervised approach is much more robust, as it does not rely on human labels and is fully explainable. The explanatory feedback can help the human to verify the decision making and potentially tune the model by selecting the position along the Pareto frontier of reward functions. We establish the timing and effectiveness of this method, demonstrating its capability for real-time performance in high-throughput and dynamic automated STEM experiments. The reward driven approach allows to construct explainable robust analysis workflows and can be generalized to a broad range of image analysis tasks in electron and scanning probe microscopy and chemical imaging.

[a] K.barakat@vols.utk.edu
[b] Sergei2@utk.edu

**Introduction**

Aberration corrected Scanning Transmission Electron Microscopy (STEM) is now the preferred technique for high-fidelity imaging, structural analysis, and spectroscopic measurements at the atomic scale[1]. STEM now routinely enables observation down to the single atom level and allows for a broad range of spectroscopic techniques including on- and off-axis Electron Energy-Loss Spectroscopy (EELS)[2,3] and diffraction imaging. Through the combined application of EELS and Z-contrast imaging[4], STEM provides detailed insights into the composition, chemical and elemental properties, such as valence state and atomic coordination, and the crystallographic structure of materials[5]. STEM-EELS has been pivotal in advancing energy research, particularly in solar cells, battery materials, and optimizing catalysts for efficient chemical processes. It has also provided crucial insights into plasmons in nano-optical structures and the study of quasiparticles and vibrational excitations, thereby enhancing our fundamental understanding in photovoltaics, sensing technologies, and solid-state physics[2,6].

The large volumes of structural and spectroscopic data routinely generated by STEM necessitates data analytics as a key part of microscope operation and generating materials-relevant physical insights. In recent years, machine learning has gained significant attention for post-acquisition data analysis, automating the extraction of meaningful patterns and features from complex datasets, uncovering structure-property relationships in multimodal spectroscopic imaging, and strengthening links to underlying physical models. One of the major classes for ML in both post-acquisition and real time experiment are the segmentation methods, using models based on deep convolutional neural networks, like U-Net[7], ResHedNet[8,9], or Mask R-CNN methods[10]. For atomically resolved image workflows, segmentation typically involves atom finding and position refinement[11-13]. Accurate atom position finding unlocks many scientifically interesting analyses, including defining defects, such as vacancies, interstitials, dislocations, and grain boundaries, and quantifying structural properties by measuring distances, angles, and coordination numbers to understand the local structural environment[14-17]. Phase identification can be performed to determine the material phase in different regions of the image.[18,19] The identified atomic positions also can be used as input for simulations to predict physical properties like electronic structure, thermal conductivity, or mechanical behavior[20,21]. Processed data can be sent to high-performance computing (HPC) environments for more complex modeling and simulations, such as molecular dynamics or density functional theory calculations[22]. Lastly, the identified structures can be used to train and improve ML models for better accuracy and efficiency in future analyses[23].

However, because these supervised methods require human labels, they are inherently limited by the labelled dataset and are sensitive to out of distribution shift effects[24]. Often, changes in imaging parameters such as magnification, sampling, and resolution severely degrade the performance of these networks. To overcome the bias and the need for human labels, semi-supervised few-shot machine learning approaches have been explored[25]. This method enables rapid classification and mapping of microstructural features in scanning transmission electron microscopy (STEM) images of oxide materials, including $SrTiO_3$/Ge heterostructures, $La_{0.8}Sr_{0.2}FeO_3$ thin films, and $MoO_3$ nanoparticles, supporting high-throughput characterization and autonomous microscopy[26]. Approaches such as ensembling[27,28], empirical risk minimization[29,30], Siamese networks[31,32], or construction of the stable representations via Barlow twin concepts[33] all can also be employed. However, both supervised and semi-supervised learning still depend on labeled data and are often associated with high latencies and high computational demands.

The newly emergent frontier is the integration of machine learning into real-time experiments, performing on the fly data analytics. In this case, the ML methods are used to assist human-based decision making by providing optimal representations of the large or high-dimensional data, facilitate immediate feedback during experiments, and enabling experiments such as precise manipulation of atomic structures[34-39]. Particularly broad range of opportunities is opened by beam-induced changes in material. While classically we aim to minimize beam damage, recently it has been shown that real-time imaging can be used for advancing our understanding of dynamic processes at the atomic scale. It allows us to observe and analyze changes such as the emergence of new defect classes in materials under beam irradiation, providing invaluable insights into phase transformations and defect dynamics. Recent examples include real-time image segmentations, using ML to identify defects of interest and use these for the spectroscopy such as EELS and high-dimensional EELS [40]. Beam induced changes can further be used to manipulate matter on the atomic level[35, 41].

Transition from post-acquisition to real-time data analysis in STEM favors the use of unsupervised learning methods, which adapt faster and can manage dynamic changes in real-time without labeled data. Many unsupervised methods also allow extension to the human-in the loop experiments, when human operator can monitor the experiment progression and tune the reward function and decision-making policies of ML agent[6, 42-44]. Recently, we have proposed the concept of reward-driven optimization workflows. In these, a sequence of simple image analysis functions is optimized within a joint parameter space to maximize physics-based rewards. These rewards can be derived from multiple considerations, such as consistency of detected structures with known physical models, human heuristics, or phenomenological laws. For cases when labeled data is available, these can be benchmarked against DCNN base methods and (implicitly) vs. human. As such, reward driven workflows offer adaptive and efficient solution, potentially overcoming the limitations of supervised deep learning methods by focusing on optimizing outcomes directly linked to the underlying physics of the system.

Here, we expand the concept of reward-driven optimization workflows to non-crystalline materials and realize this approach for real-time operation on an automated STEM. This allows us to achieve more efficient and adaptive image analysis without the need for extensive human labeling. We benchmark this approach to the classical ensembled DCNNs[45] and quantify its performance metrics.

## I. Reward driven analysis

Reward-driven workflows constructions is based on defining reward function, i.e. quantifiable success metrics that can be numerically evaluated at the end of the analysis. Once the reward function is defined, the analysis workflow, including the sequence and hyper-parameters of individual operations, can be optimized using suitable stochastic optimization frameworks, ranging from straightforward Bayesian Optimization to more multi-stage approaches such as reinforcement learning, Monte Carlo decision trees, A* and its variants, etc. [46-49]. Therefore, the main idea and objective of reward-driven workflows presented to frame the analysis as an optimization problem by formulating a reward function that can be evaluated for outputs.

In cases like atomic segmentation, rewards can be defined in a way to identify and classify all the atoms of a certain type, or all defects within the image. Here, we considered a specific task of atom finding in atomically resolved images previously applied to crystalline material[50]. Atom-finding methods, such as peak finding, Hough transforms, and Laplacian of Gaussian (LoG) approaches, require extensive parameter tuning and rely on human assessment for feedback. These simple analysis methods necessitate careful manual tuning and are often brittle, with contrast variations within a single image causing significant performance differences. We optimize the conventional Laplacian of Gaussian (LoG) algorithm, characterized by control parameters min_sigma ($\sigma_{min}$), max_sigma ($\sigma_{max}$), threshold ($Th$), and overlap ($\theta$)[51]. The parameter space consists of these variables, guiding the algorithm's tuning. Note that optimization space can also include the parameters of preprocessing space, including filtering.

The Quality Count ($QC$) function assesses the normalized difference between the number of atoms detected by the Laplacian of Gaussian (LoG) method and the expected number based on the image size and lattice parameters. It is defined as $QC = |N_{found} - N_{expected}|/N_{expected}$, where $N_{found}$ is the number of atoms detected by LoG, and $N_{expected}$ is the expected number of atoms determined by image size and lattice structure. The Error function ($ER$) evaluates the proportion of atoms that are inaccurately positioned with respect to the material's lattice structure. It is formally expressed as $ER = \#\ of\ atoms\ less\ that\ a\ threshold/N_{expected}$, where the numerator refers to the number of atoms located within a distance smaller than the specified threshold, as determined by the lattice parameters. Using multi-objective optimization, both $QC$ and $ER$ are minimized jointly, ensuring accurate atom detection and physically plausible positioning[50].

Here we implemented this already proven to be effective for crystalline materials for non-crystalline materials, benchmark it against the ELIT DCNN for validation, and demonstrates the effectiveness of the LoG* algorithm in analyzing non-crystalline structures to provides robust solutions for real-time image analysis in STEM.

## II. Benchmarking LoG*

We evaluate the effectiveness of the (LoG*) workflow in analyzing amorphous materials and materials with defects and holes using pre-acquired (STEM) data. The presence of these defects renders the reward functions approximate, since e.g. the total number of atoms now can be smaller than expected based on the image size. The performance of the LoG* workflow is compared to DCNN with the data enhanced with added noise to simulate real-world imaging conditions. This comparative analysis focuses on the fidelity of atomic positions identification and the robustness of each method under varying noise levels, providing a detailed assessment of their capabilities in handling complex material structures. To quantify the atomic finding fidelity, key performance metrics including true positive ($TP$), False Positive ($FP$), and False Negative ($FN$) predictions are defined. The ($TP$) measures the number of predicted points that are within a defined threshold distance of any actual points (Correct predictions), ($FP$) as predicted points that are not within the threshold distance of any actual points (Over-predictions) and ($FN$) measures the portion of actual points that do not have any predicted points within the threshold distance (Missed actual points). These metrics help us understand how well the model distinguishes between true, false, and missing predictions. Note that we do not explore here precise (sub-pixel) atomic position determination, since this task is trivial if the coarse position is known.

As a model system, we investigated Si-containing graphene as a material with disrupted crystallinity, where e-beam irradiation induces a variety of structural changes[52-58]. These include the formation of point and extended defects, the migration of Si atoms, the subsequent aggregation into defect clusters, and the progressive degradation of the graphene lattice. We have also used $SrTiO_3$ as a second example to compare algorithm performance for crystalline and disordered systems.

The combined visual and quantitative analysis in Figure 1(A and B) shows predicted atomic positions overlaid on the SrTiO$_3$ samples for DCNN and LoG*, respectively, across noise levels from 0.0 to 0.9. Initially, DCNN performs well at low noise levels (0.0 to 0.3), accurately detecting features with minimal false positives. However, as noise increases to moderate and high levels (0.4 to 0.9), DCNN's performance declines sharply, evidenced by a significant increase in false positives. In contrast, LoG* maintains a more stable performance across all noise levels. Although LoG* also experiences a degradation in performance with increasing noise, its false positives rise more moderately, and the true positive rate remains steadier compared to DCNN, as depicted in Figure 1(C and D). This comparative analysis highlights that while DCNN excels in low-noise environments, providing higher accuracy, LoG* demonstrates better resilience in noisier conditions, maintaining a more consistent detection rate and fewer false positives.

Same analysis has been performed on graphene sample to show the effectiveness of both methods on a non-crystalline structure. From the visual inspections in Figures 2 (A and C), it is evident that both methods at low noise levels (0.0 to 0.2), both methods perform accurately, with DCNN showing high accuracy and minimal false positives. As noise levels increase to moderate (0.3 to 0.6), Both methods begin to decline in accuracy, evidenced by more false positives and missed detections. At higher noise levels (0.7 to 0.9), although the number of false positives and false negative continues to rise for both methods, indicating challenges in distinguishing noise from actual features, LoG* shows a consistent but lower true positive prediction, suggesting better resilience to extreme noise.

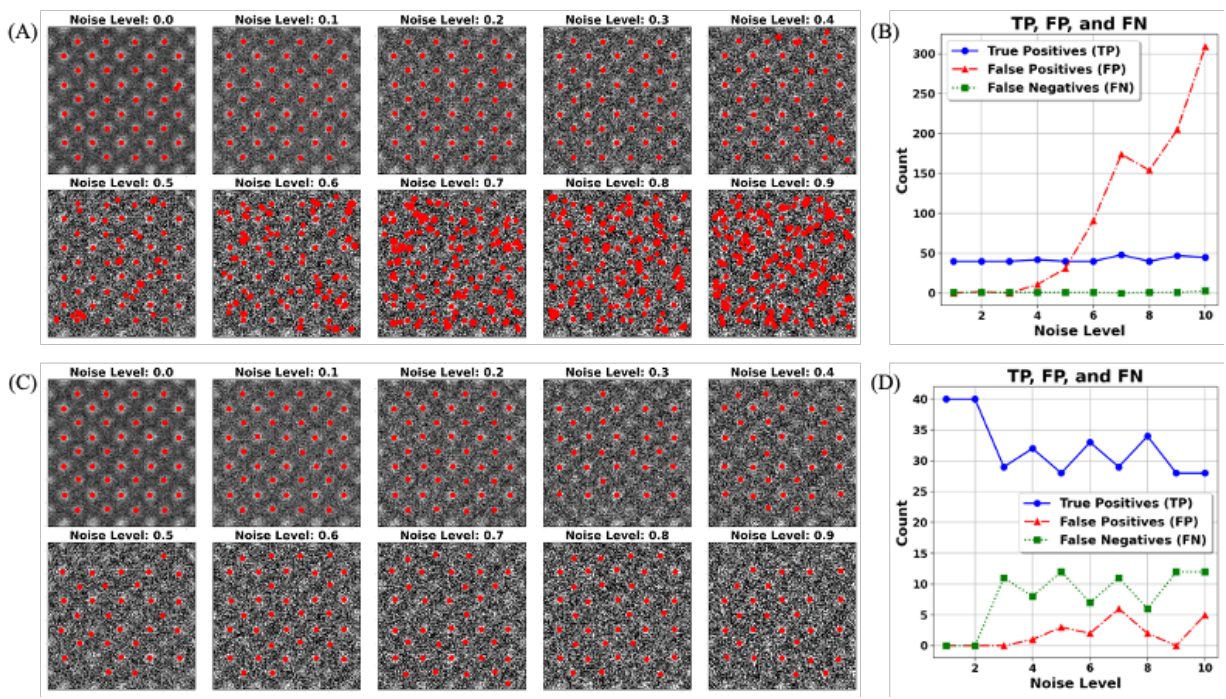

**Figure 1:** Performance comparison of LoG and DCNN on SrTiO$_3$ sample images with varying noise levels. A) Visualization of A) DCNN, C) and LoG* predicted points on Graphene sample images across different noise levels, True Positive, False Negative, and False positive predictions for B) DCNN, and D) LoG* across different noise levels.

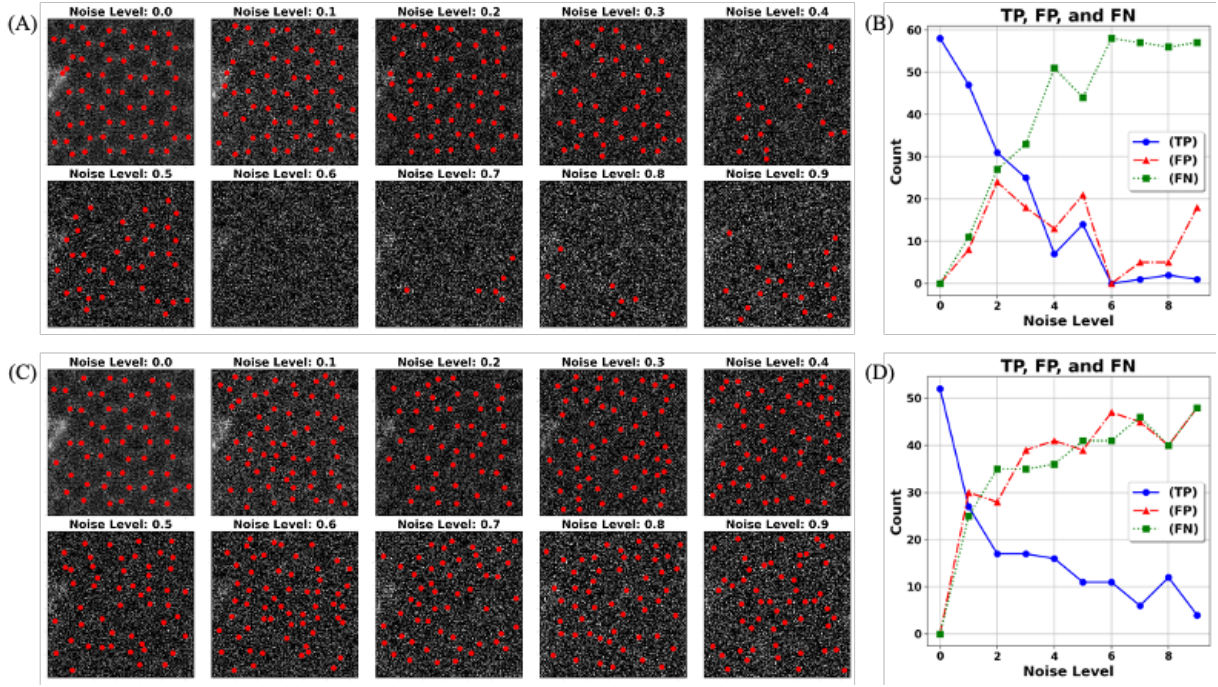

**Figure 2:** Performance comparison of LoG and DCNN on Graphene sample images with varying noise levels. A) Visualization of A) DCNN, C) and LoG* predicted points on Graphene sample images across different noise levels, True Positive, False Negative, and False positive predictions for B) DCNN, and D) LoG* across different noise levels.

This benchmarking analysis highlights that DCNN is highly effective in low-noise environments, particularly with crystalline materials where its precision can be fully utilized. However, its performance is significantly impacted by noise, which limits its applicability in more challenging environments. On the other hand, LoG* offers a more balanced performance across both crystalline and non-crystalline materials, showing greater resilience to noise and a more stable detection rate. This suggests that LoG* can be a versatile tool for real-time analysis, especially in scenarios where noise and variability in data quality are major concerns.

**III. Effects of Sliding Window on Optimized Parameter Map**

In real-time analysis, the ability to quickly process and interpret data is critical, and latencies in data analysis can hinder decision-making processes. Like neural networks and other supervised machine learning methods LoG* also needs time for optimization and optimizing the hyperparameters in the workflow. To improve its speed and make it more suitable for real-time analysis, we investigated the use of sliding windows approach. In this case, by performing LoG* on smaller images, the optimum hyperparameters pack obtained in a significantly shorter amount of time and remained relatively consistent across various small window sizes. This allows us to optimize a small section of the image and apply these parameters uniformly across the entire image, thus enhancing the practicality and efficiency of real-time analysis.

To investigate this concept, as illustrated in **Figure 3 (A),** the Laplacian of Gaussian (LoG*) method was applied to a set of sub-images (36* 256*256 pixels) extracted from a larger 512*512-pixel image. Following this, the optimized hyperparameters, which yielded true positive predictions, were analyzed for their consistency across the sub-images. As illustrated in **Figure 3 (B)**, the hyperparameters differ between sub-images, which is reasonable given the distinct content of each sub-image. This difference indicates that the model is effectively adapting to the local features of the image, tailoring the parameters based on the specific characteristics of each region. However, given the grid search configuration, where the stride was adjusted by only 20%, leading to 80% overlap between neighboring sub-images, some consistency in the hyperparameters across adjacent regions might be expected. A detailed examination of the hyperparameters, particularly $\sigma_{min}$ and $Th$, reveals a compensatory relationship. $\sigma_{min}$, which controls the minimum scale of the Gaussian kernel, tends to be lower in regions requiring sensitivity to finer details. In these cases, the threshold parameter, which determines the intensity cutoff for edge detection, is higher, to suppress background noise and prevent the detection of false positives.

The number of true positive atomic positions based on the optimum hyperparameters is relatively steady across different images within the actual image. As demonstrated in **Figure 3 (C)** some sub-images (like number 5 and 8) exhibit a significant decline in the true positive prediction of atoms due to structural defects, which adversely affects the performance of LoG*.

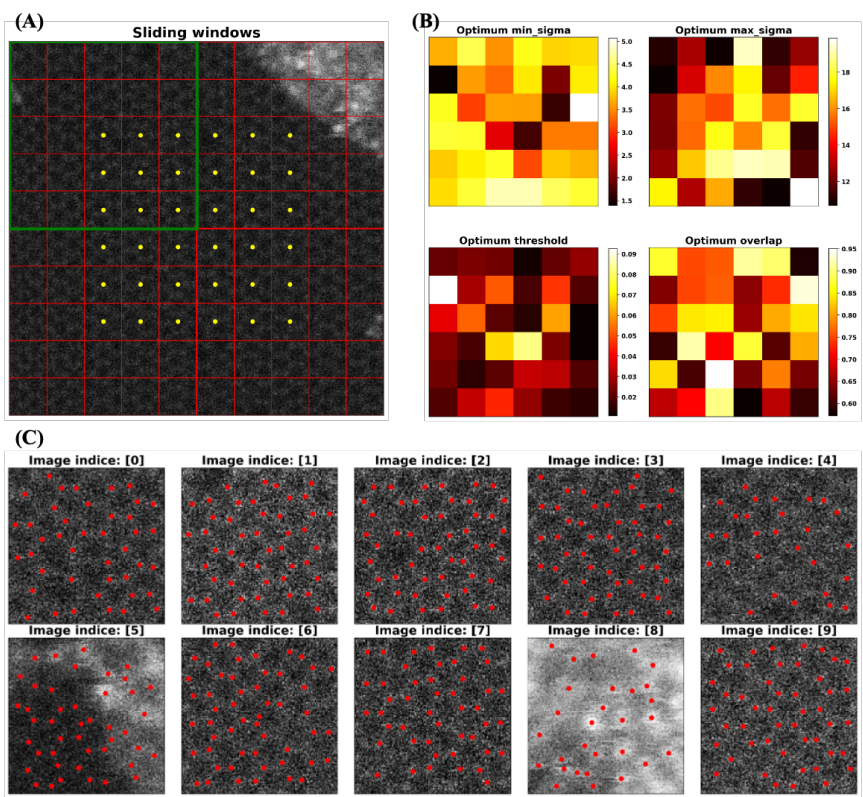

**Figure 3:** A) Grid search of 36 sub-images of 256*256-pixel, B) Optimal hyperparameters of LoG* for different locations C) LoG* prediction of true positive atomic positions on a set of ten 256*256-pixel images

The combination of the sliding window technique with the optimized parameter maps allows for a highly localized and adaptive analysis of the image. Different regions of the image are processed with parameters best suited for their unique characteristics, leading to more accurate and nuanced detection of features. This method is particularly powerful in complex images where different areas might require significantly different processing strategies to achieve optimal results.

**IV. Operation time benchmark**

As the LoG* prediction is consistent across the dataset, we further examined the timing for each prediction on a set of sub-images (10* 128*128 pixels), which the results presented in the **Figure 4** highlights distinct differences in training, optimization, and inference times of the workflows. As shown in **Figure 4 (A),** DCNN offers much faster inference, processing images in approximately 0.01 seconds per image. However, this rapid performance stems from the use of GPU computation which is necessary to handle the high computational load of the neural network architecture, and an initial investment of time and resources to create and annotate the dataset for training the DCNN model. Although this training process is a one-time effort, it can be particularly resource-intensive for large datasets. On the other hand, LoG* shows a slower inference time, around 0.1 seconds per image, due to its pixel-wise processing nature. However, **Figure 4 (B)** illustrates that when run on a CPU, DCNN does not outperform the LoG* method, underscoring the significant impact of computational resources on the efficiency of advanced neural network models. Furthermore, this performance is comparable to the data transfer rate in our system and is well above human decision-making time and hence is sufficient for many applications.

In terms of optimization/training time, as illustrated in **Figure 4(B)**, the fast inference of DCNN comes with the trade-off of a long training phase, requiring approximately 155.09 seconds. This training time is needed to construct an accurate model, particularly when dealing with large datasets. LoG*, in comparison, requires 36.25 seconds for its optimization process and bypasses the need for an extensive training phase. This enables LoG* to handle new or out-of-distribution datasets without the need for retraining. Therefore, while DCNN achieves speed during inference, LoG* offers quicker setup and greater flexibility, particularly in scenarios where data characteristics are constantly changing.

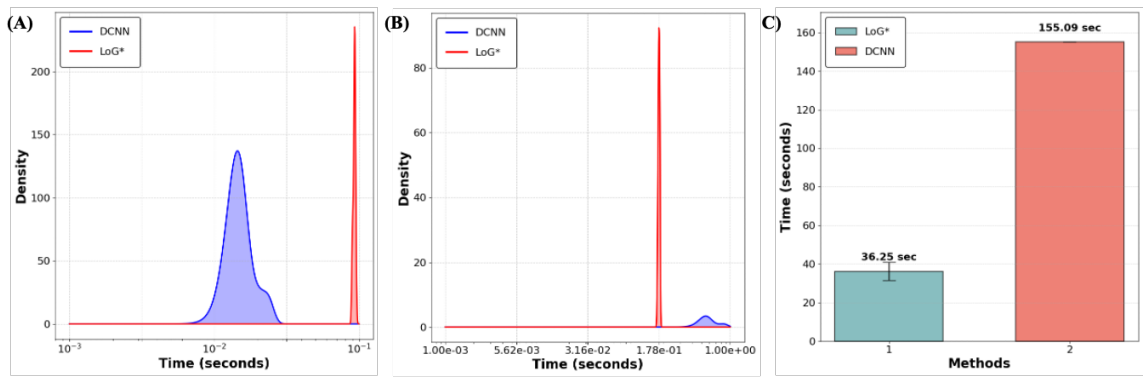

**Figure 4:** Comparative analysis of inference time for DCNN and LoG* across two scenarios: (A) using a Graphics Processing Unit (GPU) for processing a dataset of 10 images, each sized 256x256 pixels; (B) Processing the same dataset using a Central Processing Unit (CPU); and (C) Duration of DCNN training on a labeled dataset comprising 1656 images (each 256x256 pixels) relative to the average optimization time required for the LoG* method across a set of 10 images (10x256x256 pixels).

## V. Weakly supervised defect identification

The modular and transparent structure of the LoG* based reward framework facilitates the integration of both pre-processing techniques, such as noise reduction and Fast Fourier Transform (FFT) to enhance image quality, and post-processing methods like clustering to identify defects and atomic features within the target material. Following the optimization of hyperparameters and the enhancement of acquisition speed through the sliding window method, we proceed to validate the model's accuracy in identifying structural defects and atomic features. This process involves segmentation followed by clustering, which allows for the differentiation of ground truth atoms from false positives and the effective segmentation of distinct regions within the material.

As shown in Figure 5, initially, we use a sliding window method on small image patches of 128*128 pixels to determine the optimal hyperparameters for our analysis. Once these hyperparameters are identified, we apply the (LoG) to the actual image (512*512 in this case) using these parameters. Following LoG, we employ Gaussian Mixture Model (GMM) clustering[59, 60] to segment and classify similar properties and structural configurations within the material. GMM clustering is a probabilistic method that models the data as a mixture of several Gaussian distributions. In simple terms, GMM clustering assumes that the data points can be grouped into clusters, each represented by a Gaussian distribution with its own mean and variance. By fitting these Gaussian distributions to the data, GMM clustering can identify the underlying structure and assign each data point to the most likely cluster.

**Figure 5A** outlines the overall workflow, beginning with image acquisition from the microscope. The images are divided into smaller patches using a sliding window transform, where the LoG* method is applied to extract features from each patch. From this process, the optimum hyperparameters are determined, which are then applied to the full-size image. Following feature extraction, which in this case corresponds to identifying atomic coordinates, GMM clustering is employed to group these atomic features based on their similarity. **Figure 5B** illustrates the GMM clustering results, where atoms are grouped into distinct clusters, each marked by different colors to highlight variations in atomic configurations. **Figure 5D** further demonstrates this separation by displaying the clustered data in a reduced-dimensional space using Principal Component Analysis (PCA)[61, 62], emphasizing the clear distinction between atomic clusters in feature space.

To better understand both the characteristic structure and the degree of variability within each cluster, we calculated the centroids and dispersions, as shown in **Figure 5C**. The centroid heatmaps for each cluster (Cluster 0, Cluster 1, and Cluster 2) represent the average positions of the atoms, and the dispersion heatmaps show how spread out the atomic positions are around the centroids.

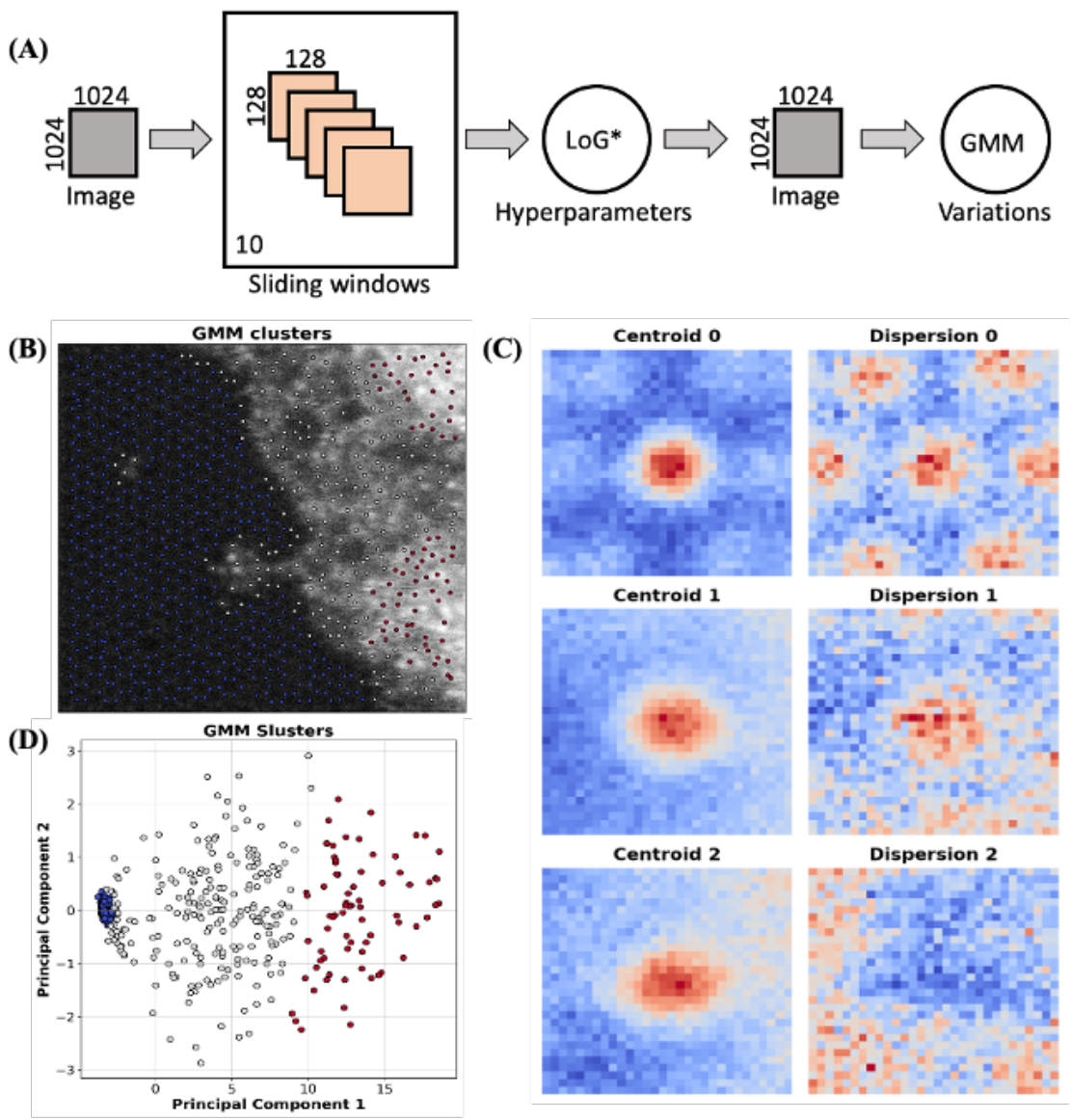

**Figure 5:** A) Designed workflow for clustering atomic types in the material structure, B) GMM clusters overlaid on the original image, with colored markers indicating atomic positions classified into different clusters, C) Centroid (left) and dispersion (right) images for each of the three GMM clusters. The centroid images represent the average features within each cluster, while the dispersion images depict the variability within each cluster, and D) Scatter plot in the space of the first two principal components, color-coded by GMM cluster assignments, showing the separation of clusters in reduced dimensionality space.

**VI. Real-time analysis and human in the loop control**

High-Angle Annular Dark Field (HAADF STEM images of graphene were captured using a monochromate, probe-corrected ThermoFisher Scientific Spectra 300 operating at 60 kV accelerating voltage with a convergence angle of 30 mrad and a nominal screen current of 100 pA. The collection angle range for the HAADF was 40-200 mrad. These images were acquired to test the proposed workflow in real time, assessing the performance and accuracy of the method under actual experimental conditions.

We present the real-time implementation of the LoG* workflow on the microscope, for various image sizes, including: 128*128, 256*256, and 512*512 pixels. For each case, LoG* was utilized to identify atomic coordinates of graphene, and its performance was systematically benchmarked against DCNN predictions. The results are illustrated in Figure 6

We also note that the reward-based workflows can be extended to the human in the loop control. Here, we utilize the fact that human operator-based analysis of imaging data is now state of the art in microscopy, and hence the key role of the real-time analysis algorithm is to empower the human to make decisions using a small number of controls. These can include both the selection of reward functions and tuning the hyperparameters of the ML algorithm.

In reward-based workflows, the hyperparameters are defined via the optimization problem. However, the presence of multiple reward functions provides a control parameter in the form of position on the Pareto front, representing the optimal balances between the objectives. This allows the flexible decision-making, enables the exploration of trade-offs, and allows to prioritize one objective over another depending on the specific needs of your experiment or system. As shown in **Figure 6**, two distinct sets of solutions were selected: one optimized for Objective 1 (minimizing error, which corresponds to reducing false positive predictions), and the other for Objective 2 (enhancing quality, reflected by an increase in true positive predictions). In this experiment, LoG* workflow consistently demonstrated reliable and precise predictions across image scales and structural complexities.

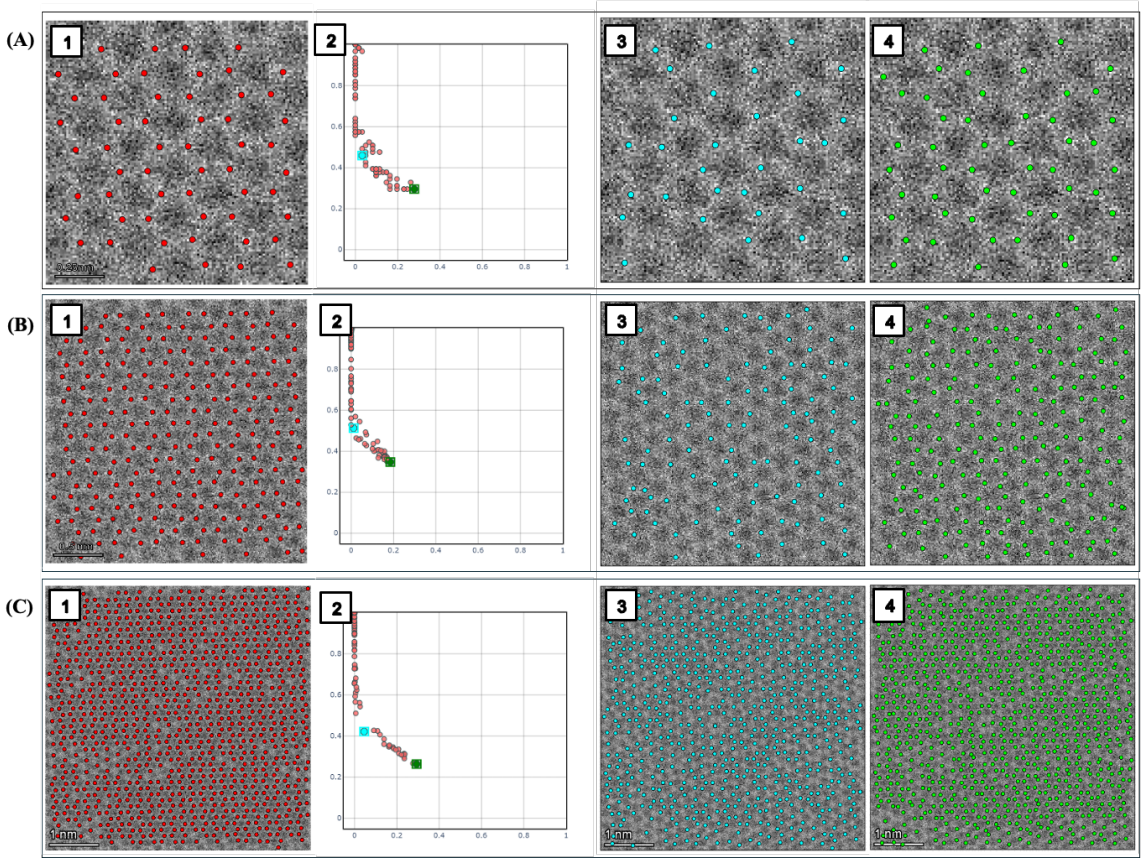

**Figure 6**: Real-time performance of DCNN and LoG* for A) 128*128-pixel image, B) 256*256-pixel image, and C) 512*512-pixel image acquired from the microscope. (1) DCNN prediction, (2) Pareto front solutions of LoG* algorithm, (3) LoG* prediction with the selected solution inclined towards Objective 1, and (4) LoG* prediction with the selected solution inclined towards Objective 2.

**VII. Summary**

We introduced the reward-driven, unsupervised workflow for real-time image segmentation in STEM, offering a probable alternative to supervised methods that rely on human-labeled data. The workflow optimizes hyperparameters of LoG as the atom finding method using physics-based reward functions, ensuring accurate atomic segmentation through multi-objective Bayesian optimization. Benchmarking of the LoG* algorithm against deep learning approaches showed that LoG* maintains consistent performance across varying noise levels and structural complexities. To enhance computational efficiency, a sliding window approach was introduced, enabling local optimization of parameters, and reducing processing time while maintaining accuracy. The operation time analysis further demonstrated that while LoG* has slower inference times than DCNNs, its faster optimization process and flexibility in handling new data make it well-suited for dynamic, real-time applications. The modular design of LoG* allows easy integration of pre- processing and post-processing techniques, such as noise reduction and clustering, for effective weakly supervised defect identification.

We have further extended the reward-based concept to the human in the loop operation. Here, selecting Pareto front solutions enables dynamic decision-making by balancing competing objectives, making it a powerful tool for high-throughput, real-time STEM analysis. Combined with the weakly supervised classification methods, this reward driven workflows can be natively deployed as a part of real-time microscopy.

Finally, we note that the key advantage of reward driven workflows is that they are explainable, unsupervised, and robust. While the range of the possible image analysis operations is more limited, this favorably compares them to the DCNN based approaches. Furthermore, many classical image analysis operations such as edge and texture detections are based on optimization towards heuristic targets, and hence can be extended towards physics-based reward concept.

**ACKNOWLEDGMENTS**:

This work (workflow development, reward-driven concept) was supported (K.B., S.V.K.) by the U.S. Department of Energy, Office of Science, Office of Basic Energy Sciences as part of the Energy Frontier Research Centers program: CSSAS-The Center for the Science of Synthesis Across Scales under award number DE-SC0019288. The work was partially supported (U.P.) by AI Tennessee Initiative at University of Tennessee Knoxville (UTK). This work was supported (A.C.H., G.D.) by the U.S. Department of Energy, Office of Science, Basic Energy Sciences, Materials Sciences and Engineering Division. The authors would like to gratefully acknowledge Dr. Ondrej Dyck of the Center for Nanophase Materials Sciences, Oak Ridge National Laboratory, for making the open data set on graphene evolution available for testing and benchmarking the LoG* prior to deployment.

**AUTHOR DECLARATIONS**

Conflict of Interest: The authors have no conflicts to disclose.

**Author Contributions:**

Kamyar Barakati: Conceptualization (equal), Data curation (lead), Formal analysis (equal), Writing – original draft (equal), Software (equal), Methodology (equal); Sergei V. Kalinin: Conceptualization (equal), Formal analysis (equal), Funding acquisition (equal), Writing – review & editing (equal), Supervision (equal); Utkarsh Pratiush: Software (equal), Writing – review & editing (equal), Investigation (equal); Austin Houston: Software (equal), Writing – review & editing (equal), Investigation (equal), Gerd Duscher: Supervision (equal), Writing – review & editing (equal), Investigation (equal).

**DATA AVAILABILITY:**

The code supporting the findings of this study is publicly accessible on GitHub at [**GitHub**]

**Materials and methods:**

**Materials:** Graphene was grown on Cu foil using atmospheric pressure chemical vapor deposition (AP-CVD). To protect the graphene and provide mechanical stability during handling, poly (methyl methacrylate) (PMMA) was spin-coated over the surface. The Cu foil was etched away using ammonium persulfate dissolved in deionized (DI) water. The remaining PMMA/graphene stack was rinsed in DI water, placed on a TEM grid, and baked on a hot plate at 150°C for 15 minutes to promote adhesion between the graphene and the TEM grid. After cooling, PMMA was removed with acetone, and any acetone residue was removed with isopropyl alcohol. The sample was dried in air and baked in an Ar-$O_2$ atmosphere (10% $O_2$) at 500°C for 1.5 hours to eliminate residual contamination. Prior to examination in the STEM, the sample was baked in vacuum at 160°C for 8 hours.

**Instrument – HPC integration:** The remote HPC client connects to the microscope (Spectra-300 manufactured by ThermoFisher, Inc) server in 40 milliseconds. Using this client, we acquire High-Angle Annular Dark Field (HAADF) images with an exposure time of 32 microseconds per pixel, producing a 512x512 pixel image. The entire process, including acquisition and data transfer to the remote PC, takes approximately 9 seconds. This acquisition process is repeated for additional images.

**DCNN implementation:** For Deep Convolutional Neural Network (DCNN) based analysis, we employ a U-Net model on the acquired image in real time. During inference, this model processes a 128x128 image in 0.0346 seconds and a 512x512 image in 0.0372 seconds to identify atoms. The DCNN is executed using an Nvidia V100 GPU.